\documentclass{ws-mpla}

\usepackage{amssymb} 
\usepackage[normalem]{ulem} 
\usepackage[dvips]{color} 

\newcommand{\txt}{\textstyle} \newcommand{\half} {{\txt \frac{1}{2}}}

\newcommand{\be}{\begin{equation}} \newcommand{\ee}{\end{equation}}
\newcommand\beq{\begin{eqnarray}} \newcommand\eeq{\end{eqnarray}}
\newcommand{\lsim}{\, \, \raisebox{-0.8ex}{$\stackrel{\textstyle
<}{\sim}$ }}  \newcommand\Mo{$\mathrm{M}_{\odot}$}
\begin{document} 
 
\markboth{Jaikumar, Reddy \& Steiner} {Quark matter in neutron stars}
 
\catchline{}{}{}{}{}
 
\title{{\bf QUARK MATTER IN NEUTRON STARS:} \\ {\bf AN APER\c{C}U}}
 
\date{today} \author{\footnotesize PRASHANTH JAIKUMAR}
\address{Physics Division, Argonne National Laboratory\\ Argonne, IL
60439 USA\\ jaikumar@phy.anl.gov}
 
\author{SANJAY REDDY AND ANDREW W. STEINER}
 
\address{Theoretical Division, Los Alamos National Laboratory\\ Los
Alamos, NM 87545 USA\\ reddy@lanl.gov, asteiner@lanl.gov}
 
\maketitle
 
\pub{Received (Day Month Year)}{Revised (Day Month Year)}
 
\begin{abstract} 
The existence of deconfined quark matter in the superdense interior of
neutron stars is a key question that has drawn considerable attention
over the past few decades. Quark matter can comprise an arbitrary
fraction of the star, from 0 for a pure neutron star to 1 for a pure
quark star, depending on the equation of state of matter at high
density. From an astrophysical viewpoint, these two extreme cases are
generally expected to manifest different observational signatures. An
intermediate fraction implies a hybrid star, where the interior
consists of mixed or homogeneous phases of quark and nuclear matter,
depending on surface and Coulomb energy costs, as well as other finite
size and screening effects. In this brief review article, we discuss
what we can deduce about quark matter in neutron stars in light of
recent exciting developments in neutron star observations. We state
the theoretical ideas underlying the equation of state of dense quark
matter, including color superconducting quark matter. We also
highlight recent advances stemming from re-examination of an old
paradigm for the surface structure of quark stars and discuss possible
evolutionary scenarios from neutron stars to quark stars, with
emphasis on astrophysical observations.

\end{abstract} 
 
\ccode{PACS: 97.60.Jd, 26.60.+c, 97.60.Gb}
 
\section{Introduction}	 
In the aftermath of a core-collapse supernova of massive stars 8 solar
masses (8~\Mo) and above, the iron core of the progenitor star
implodes from a radius of $\sim 1000$~km to a compact object of radius
$\sim 15$~km that is neutron-rich and bound by gravitational forces. A
neutron star (NS) is thus born. In 1934, Baade and Zwicky\cite{Baade}
hypothesized the supernova-neutron star connection, but did not
specify any observational signals associated to the formation of a
neutron star. Thirty years later, Bell and Hewish\cite{Bell} made the
serendipitous discovery of the first radio pulsar, interpreted
shortly thereafter by Gold\cite{Gold} as a rotating neutron
star. Since then, over 1500 similar ``rotation-powered'' neutron stars
have been identified\cite{Manchester}, with spin periods ranging from
few milliseconds to several seconds. In addition, less common
categories such as the anomalous X-ray pulsars (AXPs), soft
gamma-repeaters (SGRs) and X-ray dim isolated neutron stars (XDINS)
now exist and their relation to the ``garden-variety'' pulsars is not
well-understood. In some cases, by monitoring the pulsed and thermal
radiation from these diverse sources, important physical quantities
such as mass, radius, spin-period, magnetic fields, age and surface
black-body temperature for the neutron star can be inferred.

The main aim of neutron star observations is to understand the
underlying structure of the neutron star and its observed thermal and
magnetospheric emissions. From a theoretical standpoint, the study of
neutron star structure involves understanding the nature of matter
over an enormous density range from $\rho\sim 10^{2}-10^{15}$
g/cm$^3$. The neutron star {\it atmosphere} and {\it envelope}, having
density $\rho\leq 10^4$ g/cm$^3$ constitute a tiny fraction of the
star's volume, but can cause significant deviation from the emergent
Planckian photon flux depending on the atmospheric composition and
magnetic field strength. The {\it outer crust} of the neutron star,
from $\rho\sim 10^{4}-10^{11}$g/cm$^3$, is composed chiefly of
increasingly neutron-rich nuclei starting with $^{56}$Fe, that are
embedded in an ordered lattice structure (at zero temperature) that
minimizes the Coulomb interaction energy. The presence of free
electrons that form a degenerate Fermi sea ensures charge neutrality
and stability against $\beta$-decay. In the {\it inner crust},
somewhat above $\rho\sim 10^{11}$g/cm$^3$, neutrons drip out of
nuclei, leading to a mixture of the ordered phase with a degenerate
gas of neutrons. As densities approach the saturation density
($\rho_0\sim 2.6\times 10^{14}$g/cm$^3$) of nuclear matter, the
fraction of free neutrons increases while the nuclei may form extended
non-spherical shapes leading to rod/slab-like ``pasta'' phases (2-D
voids) and ``Swiss cheese'' phases (3-D voids). The {\it outer core}
of the neutron star, from $\rho\sim 0.5-2\rho_0$, consists of an
admixture of neutrons, protons and electrons in accordance with charge
neutrality and $\beta$-equilibirum. Other negatively charged leptonic
or hadronic species can appear when their ground state energy becomes
lower than their respective chemical potential dictated by
$\beta$-equilibrium. In the {\it inner core} ($\rho\geq 2\rho_0$),
with increasing density, negatively charged bosons can form
Bose-condensates and quarks may be deconfined, resulting in a free
quark-gluon phase where the quark pairs may also condense. These ideas
are summarized in several existing
reviews\cite{Heiselberg,Page06,Weber06}.  Currently, our constraints
in performing ab initio calculations of strongly interacting dense
matter severely limits accurate formulations of the equation of state
(EoS) at higher densities, leading to theoretical uncertainties that
preclude a definitive statement about the phase of matter in the
interior of neutron stars. However, models of the EoS that address
pure or mixed phases can be constructed and its predictions compared
to astrophysical observations to judge the viability of the
model. This brief review is aimed at assessing the evidence for the
existence of quark matter inside neutron stars, given the current
status of neutron star observations. We set the stage by recounting
recent progress in theoretical studies of dense quark matter.

\section{Dense Quark Matter} 

The critical question regarding quark matter in neutron stars is
whether the density in the interior of neutron stars is large enough so that
hadronic matter is deconfined and quarks become the relevant degrees
of freedom\cite{IK69,Collins75}. This question remains unanswered
because the underlying theory of strong interactions, Quantum
Chromodynamics (QCD), is still not sufficiently well understood at
neutron star densities.

\subsection{Color Superconductivity}

Recently, a lot of progress has been made in understanding QCD at
asymptotically high densities. In that regime, where pertubative
studies are reliable, quark matter is believed to be in a
color-flavor-locked (CFL) phase, characterized by quark pairing with a
completely gapped spectrum for single particle excitations. Such a
phase is an electromagnetic insulator in bulk and admits no electrons,
even when stressed by small quark masses\cite{Rajagopal00,Steiner02}.
If dense quark matter indeed exists inside neutron stars, where
densities are well above nuclear matter density but well below the
density where perturbative QCD is expected to be valid, the ground
state of quark matter is uncertain\cite{ABPR,Steiner05}. Nevertheless,
in such a ``hybrid star'', attractive interactions between quarks will
lead to the formation of a color superconducting
state\cite{Bailin84,Alford98,Rapp98}, characterized by quark pairing
and superfluidity. The singlet pairing gaps could be as large as 100
MeV and transport properties are strongly modified by the presence of
collective excitations below the scale of the gap.

Many phases can intervene at these intermediate densities, such as the
crystalline color superconducting phase\cite{Alford01,Bowers02}, where
quarks with different Fermi surfaces pair at non-zero momentum,
resulting in an inhomogenous but spatially periodic order parameter.
This phase spontaneously breaks translation and rotational symmetries,
and the free energy of the system is minimized when the gap varies
spatially in accordance with the residual discrete symmetries of this
phase.  Interestingly, the crystalline structure may also serve as
sites for pinning rotational vortices formed in the superfluid as a
result of stellar rotation, and could generate the observed glitch
phenomena in neutron star spin-down.

\subsection{Strange Quark Stars}

 The natural domain of physical applicability for color
superconductivity is the dense interior of neutron stars, where quark
matter may exist in a deconfined state. How would such a state arise?
Over thirty years ago, it was conjectured that at sufficiently high
density, macroscopic quark matter composed only of up and down ($u,d$)
quarks might be stabilized by the introduction of strange ($s$)
quarks, and constitute the true ground state of matter, as it would be
more bound than nuclear matter\cite{Bodmer71,Witten,FJ}. This
conjecture has not yet been decisively ruled out by experiment or
observation. The introduction of strangeness reduces Pauli repulsion
by increasing the flavor degeneracy and ensures a lower
charge-to-baryon ratio for strange quark matter compared to nuclear
matter. The latter fact can render even a large lump of strange matter
stable against fission, although it may decay by other means. On the
other hand, the stability of small lumps or ``nuggets'' of quark
matter depends on energy costs associated with surface tension and
curvature energy. In the absence of concrete results from lattice
studies of QCD at finite density and zero temperature, we rely on
simple model-dependent studies\cite{FJ} that admit a parameter window
(the parameters being the strange quark mass, the strong coupling
constant and a phenomenological Bag constant) within which bulk
strange quark matter is stable, even at zero pressure. This implies
that, if central densities inside neutron stars are large enough to
create two-flavor ($u,d$) quark matter, or if a small ``nugget'' of
cosmological/cosmic-ray origin is present, the entire neutron matter
inside the star will convert to strange quark matter by absorbing
neutrons and equilibrating strangeness. If temperatures are below the
critical temperature for color superconductivity, the strange quark
star will be a giant color superconductor.

\subsection{Nature of the Crust}

If strange quark matter is not absolutely stable at zero pressure,
then the crust of neutron stars that contain quark matter in their
core consists of hadronic matter.  The gross spectral features of such
hybrid stars tend to be very similar to neutron stars which do not
contain quark matter. Hybrid stars can be somewhat cooler than normal
neutron stars of the same age, since quark matter opens up the direct
Urca process for rapid cooling, but this effect is suppressed by quark
pairing\cite{PAC06}. On the other hand, if strange quark matter is
absolutely stable at zero pressure, then there are three possibilities
for the nature of the crust\footnote{We are referring here to the
``quark'' crust as opposed to the nuclear crust discussed in other
works\cite{Stejner}.} of the quark star.

1)\underline{{\it CFL stars:}} If the CFL phase is the ground state at
zero pressure, then the CFL phase is likely the ground state at all
densities. CFL matter extends to the star's surface and there is no crust
in the traditional sense. There is no known mechanism for pulsar glitches in
a CFL star, so not all neutron stars can be CFL stars. Further, pure
CFL stars are likely unstable to gravitational r-modes\cite{Madsen00}.

If the CFL phase is {\it not} the ground state of matter at the
surface of the strange quark star, then quark matter is positively
charged, and requires electrons to make the system charge neutral. As
explained below, this leads to two interesting possibilities for the
nature of the crust.

2)\underline{{\it The traditional paradigm:}} At the surface, positively
charged quark matter is compensated by a thin layer of electrons,
termed the ``electrosphere'', which is integrated to the quark surface
by Coulomb forces.  Solving the Poisson equation in the plane-parallel
approximation yields the profile for the electrostatic potential in
the electrosphere. In natural units $\hbar=c=1$, the profile just
outside the star's surface ($z>0$) is given (at low temperatures
$T<10^{10}$K) by\cite{UCH}

\begin{equation}
\phi=\frac{\phi_0}{(1+z/z_0)}\,,\quad
z_0=\frac{\pi\sqrt{6}}{e^2\phi_0} =501.3 \biggl(\frac{30~{\rm
MeV}}{~\phi_0}\biggr)~{\rm fm}\,,\label{eprofile}
\end{equation}

where $\phi_0$ is determined by the discontinuity in the electric
field (net charge density) at the surface.  The large electric field
that binds these electrons to quark matter leads to the Schwinger
instability of the vacuum\cite{Schwinger}, resulting in
electron-positron pair-emission, which can be an additional source of
photons in the electrosphere, which normally radiates photons via
2$\rightarrow$3 processes of Quantum Electrodynamics (QED). Therefore,
the light curves of quark stars, determined by this surface photon
emission, should be very different from neutron stars, which have
vanishing electric fields at surface.

Photon cooling calculations of bare quark stars including these
electrospheric effects as well as color superconductivity (which can
alter the specific heat and thermal conductivity of quark matter) have
been performed\cite{Page02} and are being investigated further. The
conclusion is that bare quark stars will display Super-Eddington
photon luminosities at surface temperatures $T>10^{9}$K, with a
hard spectrum that distinguishes it from thermally
radiating neutron stars.  At lower temperatures $6\times
10^8$K$<T<10^9$K, the bulk of the luminosity comes from
electron-positron pairs which subsequently imprint a wide annihilation
line on the non-thermal spectrum. Thermal emission is much suppressed
owing to plasma frequency effects in quark matter and in the
electrosphere\cite{CH}. Further cooling to temperatures $T~\sim 10^8$K
results in a non-thermal spectrum dominated by bremsstrahlung photons
from electron-electron collisions in the electrosphere\cite{JPPG}. At
temperatures just below $T<10^8$K, 2$\rightarrow$3 QED processes in
the electrosphere dominate resulting in a thermal spectrum, even
though radiation from the underlying quark matter is cutoff for
frequencies below the plasma frequency of dense quark matter
($\omega_p~\sim 20$ MeV). At very low temperatures $T\ll 10^8$K, the
luminosity of the electrosphere is exponentially suppressed.

3)\underline{{\it A new picture:}} This picture of the quark star
surface, involving homogenous quark matter and electrons, has been
recently challenged\cite{Jaikumar:2005ne}. Matter may satisfy charge
neutrality globally rather than locally, provided surface and Coulomb
costs are not prohibitively large in a heterogenous mixed phase. In
effect, relaxing the condition of local charge neutrality provides
freedom to reduce the strangeness fraction in quark matter and thereby
lower its free energy. This mixed phase would then be qualitatively
similar to the mixed phase of nuclei and electrons in the crust of
normal neutron stars and would share several features with the mixed
phase of quark drops and nuclear matter in hybrid
stars\cite{Glendenning:1992vb}.

To understand this mixed phase, we note that since the electron
chemical potential, $\mu_e$, is significantly smaller than the quark
chemical potential $\mu$ for all known models of quark matter, a
general parameterization of the EoS can be obtained by expanding in
powers of $\mu_e/\mu$ \cite{Jaikumar:2005ne},
\begin{equation}
p_{\rm QM}=p_0(\mu,m_s)-n_Q(\mu,m_s)\mu_e + \half\chi_Q(\mu,m_s)
\mu_e^2+\ldots
\label{generic_EoS}
\end{equation}
where $p_0$, $n_Q$, and $\chi_Q$ are well-defined and calculable
functions of $\mu$ and the strange quark mass, $m_s$. This
second-order expansion, which neglects the electron pressure
$p_e\sim\mu_e^4$, can be used for any model EoS or for that predicted
by QCD.

The structure of droplets in the crust of a strange quark star can be
obtained from the Poisson equation. At zero temperature and pressure,
the Gibbs free energy per quark for droplets can be compared with the
Gibbs free energy per quark for homogeneous matter. If the surface
tension, i.e. the energy cost of creating a droplet surface, is small
enough, then the crustal phase is preferred over homogeneous quark
matter. The critical surface tension is\cite{Alford06}
\begin{equation} 
\sigma_{\rm crit} = \frac{0.8 n_Q^2}{12\sqrt{\pi\alpha}\chi_Q^{3/2}} \
.
\label{crude_estimate}
\end{equation}

In the context of the Bag model for dense quark matter, the condition
for forming a mixed phase becomes
\begin{eqnarray}
\sigma \lsim 12 ~\left( \frac{m_s}{150 ~{\rm
MeV}}\right)^3~\frac{m_s}{\mu}~ {\rm MeV/fm}^2\,.
\label{eqn:sigma_bag}
\end{eqnarray}

Using two estimates of the surface energy of strangelets:(i) $\sigma
\simeq 8 $ MeV/fm$^2$ for $m_s=150$ MeV and $\mu\simeq 300$ MeV; and
(ii) $\sigma \simeq 5 $ MeV/fm$^2$ for $m_s=200$ MeV at $\mu\simeq
300$ MeV, the condition in Eq.~\ref{eqn:sigma_bag} implies that a
homogeneous phase is marginally favored for $m_s=150$ MeV while the
structured mixed phase is favored for $m_s=200$ MeV.  The sensitivity
to $m_s$ in Eq.~\ref{eqn:sigma_bag} and uncertainty in other finite
size effects can alter these quantitative estimates. If the structured
phase is favored, it will be composed of quark nuggets immersed in a
sea of electrons. The size of the quark nuggets in this phase is
determined by minimizing the surface, Coulomb and other finite size
contributions to the energy.  At low temperature, this mixed phase
will be a solid with electrons contributing to the pressure while
quarks contribute to the energy density - much like the mixed phase
with electrons and nuclei in crust of a conventional neutron
star. This modified picture of the strange star surface has a much
reduced density gradient and negligible electric field unlike the old
paradigm. In the modern viewpoint, there is no need for the
electrosphere, or large photon luminosities thereof, since matter at
the surface is globally neutral. The observed photon spectrum from
such a surface will be very different than from an electrosphere.

Neutron star observations can help in distinguishing between different
equations of state (EoS), and also between differing models for the
crust as discussed above. In the following section, we explain the
importance of such observations and their potential for advancing our
knowledge of dense matter.

\section{Neutron Star Observations:} 

Neutron star observations can be broadly classified into two
categories: (i) those that provide information about the structural
aspects of the star such as its mass and radius ; and (ii) those that
provide information about transport and cooling
processes. Observations of mass and radius provide constraints on the
EoS which typically involves physics at the energy scale set by the
baryon chemical potential ($\mu_B\simeq1$ GeV). In contrast, transport
phenomena probe the low energy response properties of the dense
interior at an energy scale set by the temperature ($T\simeq$
keV-MeV). This complementarity proves crucial in inferring the phase
structure of matter residing in neutron stars. We discuss these in
some detail below, and also address the role of transient phenomena in
determining the crustal parameters of neutron stars.

\subsection{Neutron Star Structure: Masses and Radii} 

The observation of orbital parameters in close binary systems
containing neutron stars is the classic and by far the most accurate
method of determining the component masses. Keplerian parameters
typically determine only the reduced mass of the binary system. To
infer the individual masses, general relativistic effects which lead
to post-Keplerian corrections to the orbital evolution need to be
measured.  There are five known types of binary systems containing
neutron stars: (1) double neutron star systems containing a pulsar
(PSR) and a neutron star; (2) neutron star-white dwarf systems
containing a PSR and a white dwarf (WD); (3) High mass X-ray binaries
(HMXBs) containing a neutron star and a massive companion star ($M >
10 M_\odot$); (4) Low mass X-ray binaries (LMXBs) consisting of a
neutron star and a companion star with mass $M <1 M_\odot$; and (5)
neutron star-black hole (BH) binaries. In compact binary systems, such
as double-neutron star and PSR-WD systems, orbital decay due to
gravitational wave-emission, advance of periastron and Shapiro
time-delay have been measured (for an excellent review of general
relativistic orbital effects in compact binaries see\cite{Stairs}).

Once the post-Keplerian parameters are measured, it overdetermines the
set of quantities needed to infer the individual masses and thereby
allows a high precision neutron star mass measurement virtually free
of any systematic error. In these systems, the neutron star masses lie
in the range 1.18-1.44 $M_\odot$ and have errors of less than a tenth
of a percent. This tight clustering of masses at a relatively low
value $\sim 1.4 M_\odot$ warrants an explanation. One possible
explanation proposed by Bethe and Brown\cite{Brown} is that the
maximum mass of a neutron star is $\simeq 1.5~M_\odot$; any heavier
and they would become black-holes. A more mundane explanation is based
on the evolution of HMXBs which are suspected to be the progenitors of
double neutron star systems. The lifetime of HMXBs is short because
the massive star evolves rapidly on a time scale of a few million
years - too short for significant accretion from the massive star to
increase the mass of the neutron star significantly from its mass at
birth, which is expected to lie in the range 1.2-1.5 $M_\odot$.

Clearly, the discovery of a massive neutron star in systems where
adequate mass accretion is possible with a mass close to 2 $M_\odot$
would disprove the Bethe-Brown scenario and lend credence to the
evolutionary argument. PSR+WD systems are prime candidates for finding
heavy neutron stars. These systems are thought to have evolved from
LMXBs which have long lifetimes and where significant accretion is
expected to occur. Indeed a candidate heavy neutron star called PSR
J0751+1807 has been found in the NS+WD system. Measurement of the
orbital decay and Shapiro delay have yielded a neutron star mass
$M=2.1\pm0.2~M_\odot$\cite{Nice05}. This along with the possible
confirmation (at 95\% confidence) of at least one pulsar with
$M>1.68$\Mo, following the recent detection of 21 millisecond pulsars
in the globular cluster Terzan 5, 13 of which are in
binaries\cite{Ransom05}, adds to the emerging trend toward relatively
large mass neutron stars.

\begin{figure}
\includegraphics[width=0.8\textwidth,angle=-90]{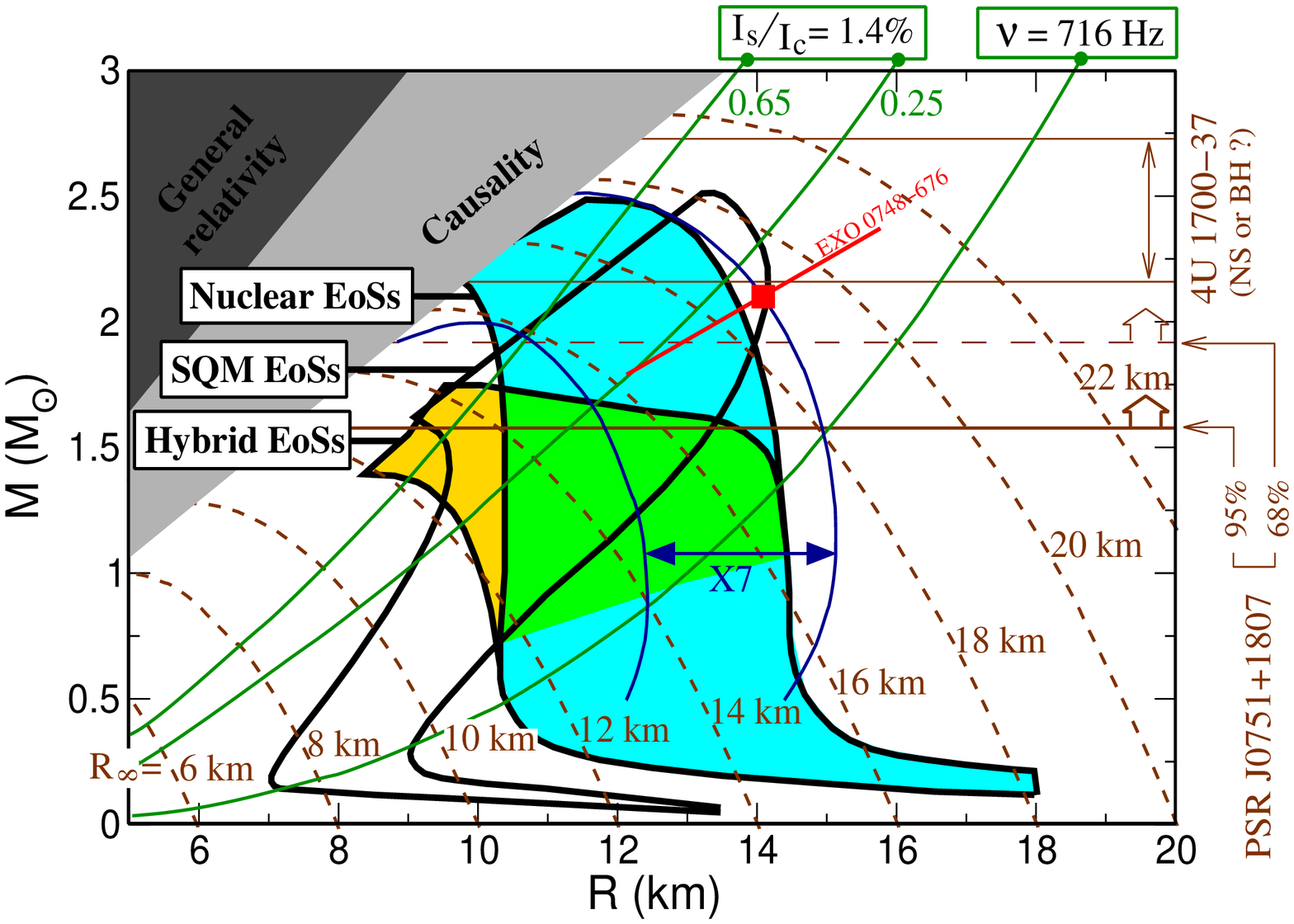}
\caption{Mass-Radius constraints from observations and model
predictions for the mass-radius of nucleonic stars, hybrid stars and
strange quark stars. Original figure adapted from Page \& Reddy
(2006)}
\label{fig:m-r}      
\end{figure}

Radius measurements of neutron stars have become even more crucial in
pinning down the equation of state at high density, now that mass
ranges have widened. In principle, determining the radius of a neutron
star may seem fairly straightforward. If the neutron star radiated
like a black body, then X-ray observations would be able to determine
both the flux $f$ and the spectral temperature $T$. Further, if the
distance to the object $d$ were known, the radius of the star $R$ can
inferred from the relation between the observed flux and the
temperature $ f=4 \pi R^2 \sigma_{SB} T^4/d^2$ where $\sigma_{SB}$ is the
Stefan-Boltzmann constant.  In practice however there are several
complications that make radius measurements a challenging task: (i)
even for isotropic black bodies, the observed flux, temperature and
apparent radius are all modified due to the effects of gravitational
red-shift and it is only possible to infer the radius at infinity
which is related to the true radius of the star through the relation
$R_\infty=(1+z)~R$ with the red-shift factor
$(1+z)^{-1}=\sqrt{1-2GM/Rc^2}$. Consequently, instead of measuring a
radius we can only infer a relation between mass and radius as shown
in Fig.\ref{fig:m-r} where curves corresponding to different values of
$R_\infty$ are depicted as dashed curves; (ii) the assumption that
neutron stars radiate isotropically with a black body spectrum is
seldom true because most neutron stars are characterized by magnetic
fields and atmospheres; (iii) it is not possible to measure the
luminosity or flux of the object directly due to inter-stellar
absorption and hence modeling the atmosphere and the role of magnetic
fields become crucial; and (iv) although the distance measurements can
in principle be obtained through parallax, accurate measurements have
been difficult to obtain.

 A promising candidate class for radius measurement is quiescent
 neutron stars in LMXBs situated in globular clusters. Their magnetic
 fields are small (B$<10^{10}$ G), and their atmosphere is very likely
 to be composed of hydrogen accreted from the companion. Since
 distances to the globular clusters are typically well-known,
 uncertainties in determining the radius are minimal. This is
 exemplified by the recent determination of $R_\infty$ in the
 quiescent LMXB called X7 in the globular cluster 47
 Tucanae\cite{Heinke}. A model hydrogen atmosphere with consistent
 surface gravity was employed to obtain the 90\% confidence contours
 in the mass-radius plane shown in Fig.\ref{fig:m-r}.

It has recently been realized that compact objects in LMXBs which
exhibit X-ray bursting behavior may provide a promising new avenue to
determine, simultaneously, both the mass and radius of a neutron
star\cite{Ozel}. In these objects, there is the potential to observe,
in addition to the quiescent luminosity that can be used to infer
$R\infty$, the Eddington luminosity during the burst and the
gravitational red-shift through direct observation of the shift in
identifiable atomic absorption lines in the atmosphere. The peak
luminosity of the burst can be identified with the Eddington
luminosity if it remains constant over time and over several bursts
(Eddington luminosity $L_{\rm Edd}=(4\pi cGM/\kappa)(1+z)$ is
essentially determined by the mass of the object and is the maximum
radiation possible in equilibrium wherein the radially outward
radiation force in the shell exactly counterbalances gravity). In her
recent article\cite{Ozel}, Ozel proposes that observations of bursting
behavior in the low-mass X-ray binary EXO 0748-676 already provide
detailed information about all three aforementioned quantities. Since
these quantities have different dependences on mass and radius their
simultaneous determination in turn allows for the independent
determination of mass (2.10$\pm$ 0.28~\Mo) and radius (13.8$\pm$1.8
km)\cite{Ozel}. This is also shown in Fig.\ref{fig:m-r}, where it has
been the assumed that the gas accreted from the companion has solar
composition\cite{Ozel}.

In Fig.\ref{fig:m-r}, the mass radius predictions for different model
EoS are shown. We have broadly classified models into three
classes/regions in the mass-radius plot: (1) Nucleonic stars; (2)
Hybrid stars characterized by a soft EoS in the interior (due to a
phase transition to an exotic state) surrounded by a nuclear shell;
and (3) Strange quark stars made entirely of stable strange quark
matter. The heavy neutron star candidates and the rather large
inferred radii disfavor the scenario in which significant softening
due to a phase transition at high density occurs. Strange stars and
possibly even hybrid stars with a fairly stiff high density EoS remain
viable.

\subsection{Thermal Evolution}

Observations of thermally emitting neutron stars provide another
handle, besides mass and radius measurements, to probe the interior of
neutron stars.  The cooling history of thermally emitting neutron
stars in the first million years of their life is governed by neutrino
emission from the dense interior. The low lying excitation spectrum of
quasiparticles and phase structure of matter plays a role in the
neutrino emission rates, enabling observational constraints on the
neutron star cooling rate to constrain the interior physics. As
mentioned previously, x-ray observations of neutron star surface
temperatures are complicated by atmosphere and magnetic field
considerations. The other key ingredient needed is the age of the
neutron star. There are typically three methods used to determine the
age: (i) direct association with a historic supernova; (ii)
association with a supernova remnant's measured expansion rate and
radius of the nebula or association with a measured neutron star
velocity and distance; and (iii) spin-down age as inferred from the
measurement of the period and period derivative assuming spin-down is
due to magnetic dipole radiation. There are three young neutron stars
which are associated with historic supernova: (1) PSR B0531+21 (the
Crab Pulsar); (2) CXO J232327.8+584842 (Cas A); and (3) PSR J0205+6449
(or 3C58). In these young systems 300-1000 years old, there are
significant sources of non-thermal emission and the thermal emission
itself is not detectable. Consequently, only upper limits on the
thermal radiation can be inferred.  As we shall discuss later these
limits already provide useful constraints. In about 10 other sources
thermal radiation from the surface has been observed. In these cases
the inferred luminosity or temperature depends on the assumptions made
about the NS atmosphere but they nonetheless provide valuable data to
constrain NS cooling at late times ($10^3-10^7$ yrs).

Through detailed modeling efforts by several groups it is possible to
relate the interior cooling directly to the surface temperature (for a
recent review see\cite{Yakovlev:2004iq,Page2004}). These studies
have shown that surface temperature depends sensitively on the
composition of the envelope in the uppermost regions of the star just
below the photon emitting region. This region can support the largest
temperature gradients due to their small thermal
conductivity\cite{Page2004}. Typically, light elements like H, He, C
or O have a larger conductivity relative to the heavier Fe-like
elements and thereby result in higher luminosity at early times. The
change in luminosity due to the compositional changes can be as large
as a factor of 10 both at early and late times. With this
observational background and caveats in mind, we now turn to a
discussion of how neutrino emission rates in different high density
phases can impact neutron star cooling.

In a broad sense, neutrino cooling in dense matter can be either fast
or slow.  Fast cooling neutrino processes are those that can occur at
the one (quasi-) particle level, while slow cooling is due to two
particle processes such as Bremsstrahlung reactions. In nucleonic
matter, the one-particle process would be $\beta$-decay of neutrons
and its inverse reaction ($n\rightarrow pe^-\bar{\nu}_e$ and
$e^-p\rightarrow n \nu_e$). This reaction, which is called the
direct-URCA (DURCA) reaction, when kinematically allowed, leads to a
rapid energy loss rate $\dot{\epsilon}_{\nu} \simeq 10^{26}~T_9^6$
ergs/cm$^3$/s for typical densities characteristic of neutron star
interiors. However, several models of dense nuclear matter predict a
relatively small proton fraction which in turn forbids the DURCA
reaction because the discrepancy between the neutron and proton Fermi
momenta is too large to satisfy momentum conservation. Under these
conditions a spectator nucleon is required to satisfy momentum
conservation. Such reactions which are similar to Bremsstrahlung
reactions lead to a significantly slower rate of energy loss
$\dot{\epsilon}_{\nu} \simeq 10^{21}~T_9^8$ ergs/cm$^3$/s. Whether or
not DURCA reactions can occur in neutron stars with nucleonic matter
remains an open issue; in particular it appears likely that even if it
is forbidden in light neutron stars with mass $\sim 1.4 M\odot$ it may
occur in heavier stars. This large range of allowed emissivities even
in the standard case is unfortunate from the point of view of
constraining novel high density phases.

In nuclear as well as quark matter, superfluidity plays a crucial role
as it suppresses conventional single-particle beta decay rates, at the
same time opening up new pathways for neutrino emission through
pair-breaking and recombination or via other correlations in novel
ground states of dense matter\cite{Flo,JP,JPS,RST}. Thus, admixtures
of completely different phases can explain the cooling history of
neutron stars equally well\cite{Page00}. For example, hybrid stars
with color superconducting quark cores and a mantle of normal nuclear
display cooling curves that are consistent with present surface
temperature observations of neutron stars. Nevertheless, at present,
the bulk of the data is consistent with standard neutron star cooling
models which have superfluid effects included but disallow DURCA. One
exception is the observation of the coldest neutron star PSR
J0205+6449 (3C58) which is only marginally consistent with cooling
curves based on these models since it relies on rather finely tuned
singlet proton pairing and weak triplet neutron pairing.  Interesting
new luminosity limits are coming from studies of supernova remnants
(SNR) reported in Ref. \cite{Kaplan2006}. These limits come from the
non-observation of thermal flux from the as yet unidentified neutron
star and are compelling only in a statistical sense since we expect a
fair number of the SNR to contain neutron stars. Nonetheless the cold
neutron star in 3C58 and the growing number of SNR with low observed
luminosity may well suggest that some neutron stars require some type
of rapid cooling mechanism.

\subsection{Transient Phenomena} Additional observed phenomena such as glitches, quasi-periodic oscillations in accreting neutron stars, thermal radiation from
quiescent LMXBs and seismic vibrations during magnetar flares (SGRs)
can potentially yield valuable information about the neutron star
interior and constrain the EoS. Glitches refer to the sudden spin-up
of pulsars that otherwise gradually spin down. Although detailed
modeling of the glitch behavior is still quite uncertain, they are
thought to arise from the catastrophic unpinning of superfluid
vortices in the neutron star crust. The observed $\Delta
\dot{\Omega}/\dot{\Omega} \simeq 10^{-3}-10^{-2}$ and the slow slow
post-glitch relaxation of the NS spin down rate favors the existence
of a superfluid component or at the very least some component that
decouples and subsequently couples to the neutron star spin. The inner
crust of the neutron star is a particularly favored location for the
glitch since here a neutron superfluid coexists with a lattice of
nuclei. Although the mechanism for the coupling to the superfluid is
not fully understood quantitatively, it offers a natural explanation
for glitch dynamics\cite{Alpar1985}.  Strange stars with homogeneous
quark matter cannot support a nuclear crust with a coexisting
superfluid. It has been argued that the very phenomenon of glitches
disfavors the strange star scenario\cite{Alpar1987}. However, recent
advances in understanding the phase structure of strange quark matter
suggest that structured ground states where a lattice and a superfluid
coexist can occur. These developments indicate that further work is
necessary before we can rule out strange stars on the basis of
glitches.

Giant flares observed in soft gamma repeaters (SGRs), now believed to
be highly magnetized neutron stars called magnetars, show
quasi-periodic oscillations (QPOs) in the tail of the
burst\cite{Watts:2005} at frequencies of few tens to few hundred
Hz. There is exciting preliminary evidence that these are seismic in
origin and are due to shear modes excited in the neutron star
crust. If confirmed, it provides a direct means to measure both the
composition and the radial extent of the crust\cite{Piro:2005}.  The
detection of a QPO at 626.5 Hz in the 2004 Hyperflare of SGR 1806-20
is of particular interest since it is likely to be an n=1 mode that is
sensitive to the the radius of the crust\cite{Watts:2005}.  If this
identification is secure it restricts the crust thickness to be
$\sim1$ km, which is much larger than the crust atop a strange star
($\sim$ 0.1 km).

\section{Evolution of neutron stars and quark stars} 

Finally, we review recent interesting suggestions on evolutionary
scenarios that can explain apparently different categories of neutron
stars.  Anomalous X-ray pulsars or AXPs, so named because their strong
X-ray emission is surprising given their low spin frequency, and Soft
gamma-ray repeaters (SGRs), are believed to be neutron stars which emit
irregular bursts of low-energy gamma rays. Could it be that these two
types of objects are actually evolving quark stars with extremely
large magnetic fields? Recent work\cite{Niebergal}, based on
magneto-hydrodynamic studies of a quark star's magnetic field, goes
further in positing an attractive evolutionary picture that connects
AXPs/SGRs with XDINS. AXPs/SGRs are conjectured to be quark stars
whose interiors are superfluid except for rotational vortices that
entrain the magnetic fields. The misalignment between the magnetic
field in the vortex and the external dipole field is removed in a
short time period following the formation of the superfluid state
($t\sim 0.1$~sec) with rapid external magnetic reconnections that
produce the energetic X-ray bursts seen from
AXPs/SGRs\cite{Dobler}. As these stars spin down rapidly due to their
large magnetic fields, magnetic field lines are expelled along with
the quantized vortices, thereby increasing the spin-period and
decreasing the spin-down rate. The evolutionary track of such stars
can explain the suggestive period clustering of XDINS, as well as the
lack of radio emissions, provided the star is sufficiently compact
($R<10$ km). However, this model cannot as yet explain the broad
absorption line seen in the spectrum of some XDINS, or the latter's
slow pulsations, and the model assumes the interior superfluid to be
an electromagnetic insulator, such as the CFL phase,
which may not be the ground state at moderately high density.

The diversity of sources identified as neutron stars indicates that
interior compositions may differ from one category to another,
although the equation of state must be unique. This leads naturally to
the question of {\it which} neutron stars are likely to contain quark
matter in their interior.  This likelihood question was addressed
recently by Staff et al.\cite{Staff}, who studied the role of
spin-down of isolated neutron stars in driving quark deconfinement in
their high density core. Assuming spin-down to be solely due to
magnetic braking, they obtained typical timescales to quark
deconfinement for neutron stars that are born with Keplerian
frequencies. The minimum and maximum neutron star masses that allow
for deconfinement (via spin-down only) were identified, based on
plausible EoS.  Their results suggest that neutron stars lighter than
$1.5$\Mo~can not reach a deconfined phase. Further, depending on the
EoS, neutron stars of more than $1.5$\Mo~can enter a quark phase only
if they are spinning faster than about 6 milliseconds as observed now,
whereas larger spin periods imply that they are either already quark
stars or will never become one. Thus, quark deconfinement is more
likely in light, rapidly spinning neutron stars, especially if the
deconfinement threshold density is low ($<5\rho_0$). In this context,
given that EXO 0748-676 has a deduced spin-period of 47Hz (uncommonly
slow for an LMXB) and its high mass, it is clearly not a quark star
and is not likely to suffer quark deconfinement in the future, unless
the deconfinement threshold is much lower than 5$\rho_0$. A low
deconfinement threshold is still not ruled out since very stiff
equations of state can reproduce the observed mass and radius of EXO
0748-676 with a central density of just
2$\rho_0$\cite{MS0}. Therefore, hybrid stars remain very much a
possibility while bare quark stars, which are based on the absolute
stability of strange quark matter at zero pressure, face serious
difficulties in explaining these observations of high mass neutron
stars. It is also worth pointing out that contrary to popular belief,
low-mass X-ray binaries (LMXBs) are {\it not} expected to contain
hybrid stars since the mass increase from accretion is more than
compensated for by the concomitant spin-up and magnetic field
quenching (which greatly increases the spin-down time to deconfinement
densities), so that central densities are actually lowered, not
raised.

\section{Summary}

It is apparent that recent neutron star observations disfavor the
appearance of soft exotic (non-hadronic) matter at high density. They
do not as yet completely rule out all quark matter equations of
state. Hybrid stars may exist only as a small population among neutron
stars and should not be dismissed until long-term evolution of neutron
stars and transient phenomena occuring at neutron star surfaces are
understood within a consistent neutron star picture. In particular, it
is premature to rule out quark matter in neutron stars on the basis of
mass-radius measurements alone.

\section*{Acknowledgments} 
The authors acknowledge discussions with Rachid Ouyed, Denis Leahy and
Kaya Mori. P.J. is supported by the Department of Energy, Office of
Nuclear Physics, Contract no. W-31-109-ENG-38. S.R. and A.W.S are
supported by the National Nuclear Security Administration of the
U.S. Department of Energy at Los Alamos National Laboratory under
Contract No. DE-AC52-06NA25396.


\bibliography{quarkstar} \bibliographystyle{aip}

\end{document}